\documentclass[prl,superscriptaddress,twocolumn]{revtex4}%
\usepackage[ansinew]{inputenc}
\usepackage{graphicx}
\usepackage{amsmath}
\usepackage{amsthm}
\usepackage{bm}
\usepackage{layout}
\usepackage{float}
\usepackage{txfonts}
\usepackage{amsfonts}
\usepackage{amssymb}%
\begin{document}

\title{Macroscopic realism and spatiotemporal continuity}

\begin{abstract}
Macroscopic realism, as introduced by Leggett and Garg, is the world view in
which properties of macroscopic systems exist independent of and are not
influenced by measurement. Motivated by classical physical laws such as
Newtonian mechanics or Maxwell's electrodynamics, in this work we add the
restrictive postulate that the observables of macroscopic objects are evolved
\textit{continuously through space and time}. Quantum theory violates both
macroscopic realism and the continuity assumption. While decoherence or
collapse models (e.g.\ due to a universal noise background or gravitational
self energy) can restore macroscopic realism, we show that a continuous
spatiotemporal description does not become possible in general. This shines
new light on the question how the classical world arises out of the quantum realm.

\end{abstract}
\date{\today}%

\author{Johannes Kofler}%
%

\affiliation
{Faculty of Physics, University of Vienna, Boltzmanngasse 5, 1090 Vienna, Austria}%
%

\affiliation
{Institute for Quantum Optics and Quantum Information, Austrian Academy of Sciences, Boltzmanngasse 3, 1090 Vienna, Austria}%
%

\author{Nikola Buri\'{c}}%
%

\affiliation{Institute of Physics, Pregrevica 118, 11080 Belgrade, Serbia}%
%

\author{{\v C}aslav Brukner}%
%

\affiliation
{Faculty of Physics, University of Vienna, Boltzmanngasse 5, 1090 Vienna, Austria}%
%

\affiliation
{Institute for Quantum Optics and Quantum Information, Austrian Academy of Sciences, Boltzmanngasse 3, 1090 Vienna, Austria}%
%

\maketitle

Classical laws, as they are formulated in mechanics or electrodynamics, (i)
give a \textit{continuous spatiotemporal evolution} of a system's properties
and (ii) also are in agreement with the theory-independent concept of
\textit{macroscopic realism} (macrorealism)~\cite{Legg1985}. For example, the
position or angular momentum of a macroscopic body evolve continuously in
space and time as governed by classical laws in the form of certain
differential equations. These physical properties are macrorealistic,
i.e.\ they can be assumed to exist prior to observation and to be not
influenced by them. The predictions of quantum mechanics violate macrorealism.
Decoherence is the mechanism by which---through interaction with an
environment---the non-diagonal terms of a system's density matrix are
suppressed, turning quantum states into statistical mixtures \cite{Zure2003}.
We show that, although decoherence can establish macrorealism, it does not
necessarily lead to a continuous spatiotemporal evolution of macroscopic variables.

Macrorealism (MR) bases on three postulates~\cite{Legg2002}:

\begin{quote}
"(1) \textit{Macrorealism per se}. A macroscopic object which has available to
it two or more macroscopically distinct states is at any given time in a
definite one of those states.

(2) \textit{Non-invasive measurability}. It is possible in principle to
determine which of these states the system is in without any effect on the
state itself or on the subsequent system dynamics.

(3) \textit{Induction}. The properties of ensembles are determined exclusively
by initial conditions (and in particular not by final conditions)."
\end{quote}

These assumptions allow to derive the so called Leggett-Garg inequality for
temporal correlations, whose violation indicates the non-classicality of a
macroscopic object~\cite{Schr1935}. Notwithstanding recent achievements that
could demonstrate quantum interference in large systems~\cite{Arnd1999}, the
high experimental demands for a demonstration of the violation of macrorealism
have not been achieved yet.

One can distinguish various levels of \textquotedblleft
classicality\textquotedblright\ for a macroscopic physical system. The most
abstract level is the postulation that systems obey the Leggett-Garg
inequality and macrorealism (MR). We now introduce the more restrictive notion
of classicality as the conjunction of macrorealism and \textit{continuity}
(MR\&C). MR\&C bases on the three postulates of MR as well as on a fourth one:

\begin{quote}
(4) \textit{Continuity}. The observables of macroscopic objects evolve
continuously through space and time.
\end{quote}

Notwithstanding MR\&C represents a narrower class of theories than MR,
continuity is a natural assumption because even the allegedly discrete and
abrupt events in the physical world like the result of a dice toss stem from a
continuous evolution of all objects through space and time.

\begin{figure}[t]
\begin{center}
\includegraphics{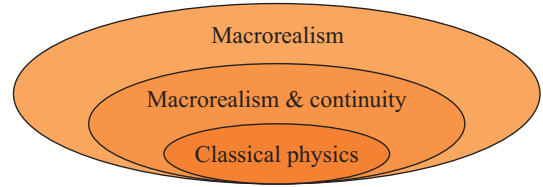}
\end{center}
\par
\vspace{-0.25cm}\caption{(Color online.) Illustration of the different levels
of classicality: "Macrorealism" (MR) in the Leggett-Garg definition,
"Macrorealism and continuity" (MR\&C) as put forward in this work, and
"Classical physics" with concrete laws of motion.}%
\label{Figure1}%
\end{figure}The most restrictive level is classical physics itself where
systems obey \textit{concrete laws} such as Newton's or Maxwell's equations.
While it is clear that validity of classical physics implies validity of MR\&C
but not the opposite, we will show that the theory sets obey the strict
relation "Classical physics$\,\subset\,$MR\&C$\,\subset\,$MR". This is
illustrated in Figure~\ref{Figure1} and has interesting consequences for the
quantum-to-classical transition, as we will present a situation where due to
environmental decoherence (or collapse models~\cite{Ghir1986}) MR is fulfilled
but MR\&C is not. Since decoherence is extremely hard to avoid, our results
should be interpreted as a chance for experiments to demonstrate a certain
level of non-classicality despite the action of decoherence.

In Refs.~\cite{Kofl2007,Kofl2008} it was demonstrated that the Leggett-Garg
inequality can always be violated under sharp measurements even for
arbitrarily large systems, using as an example a quantum spin-$j$. However,
one can speak about a violation of macrorealism only if sufficiently
coarse-grained \textquotedblleft classical\textquotedblright\ measurements are
employed which distinguish macroscopically distinct states. Under
coarse-grained measurements, one can distinguish between two types of time
evolutions. For the set of \textquotedblleft classical\textquotedblright%
\ Hamiltonians, the temporal correlations satisfy macrorealism and even obey
classical laws of motion. However, there exist \textit{\textquotedblleft
non-classical\textquotedblright\ Hamiltonians} which allow to violate
macrorealism even under coarse-grained measurements. Essentially, they need to
build up macroscopic superpositions in time. In this paper, we show that
decoherence validates MR. However, whether or not non-classical Hamiltonians
lead to a violation of MR\&C depends on the concrete model of decoherence.
Sufficiently strong thermal (dissipative) environments may establish the
validity of MR\&C, while even arbitrarily strong dephasing (non-dissipative)
environments are not able to accomplish that.

We now introduce the basic mathematical concepts for the further analysis. The
(normalized and positive) $Q$-distribution of a quantum spin-$j$ state
$\hat{\rho}$ is given by~\cite{Agar1981} $Q(\Omega)\equiv\tfrac{2j+1}{4\pi
}\,\langle\Omega|\hat{\rho}|\Omega\rangle$ with $|\Omega\rangle$ the spin
coherent states~\cite{Radc1971}. In a coarse-grained spin measurement, the
whole unit sphere is decomposed into a number of mutually disjoint angular
regions $\Omega_{k}$ where the (polar and azimuthal) angular size of these
regions, $\Delta\Theta$, has to be much larger than the inverse square root of
the spin length $j$: $\Delta\Theta\gg1/\!\sqrt{j}$. A (POVM) coarse-grained
measurement with elements $\hat{P}_{k}\equiv\tfrac{2j+1}{4\pi}\,%
{\textstyle\iint\nolimits_{\Omega_{k}}}
|\Omega\rangle\langle\Omega|\,$d$^{2}\Omega$ finds out in which of these
coarse-grained \textquotedblleft slots\textquotedblright\ the quantum system
is ($%
{\textstyle\sum\nolimits_{k}}
\hat{P}_{k}=\openone$)~\cite{Kofl2008}. The outcome $k$ is found with
probability $w_{k}=\,$Tr$[\hat{\rho}\hat{P}_{k}]$ or just via integration over
the $Q$-distribution, representing a classical ensemble of spins: $w_{k}=%
{\textstyle\iint\nolimits_{\Omega_{k}}}
Q(\Omega)\,$d$^{2}\Omega$. This reflects macrorealism per se.

Upon measurement the state $\hat{\rho}$ is reduced to $\hat{\rho}_{k}=\hat
{M}_{k}\hat{\rho}\hat{M}_{k}^{\dag}/w_{k}$, with $\hat{M}_{k}$ the Kraus
operators obeying $\hat{M}_{k}^{\dag}\hat{M}_{k}=\hat{P}_{k}$. Its
corresponding $Q$-distribution is $Q_{k}=\tfrac{2j+1}{4\pi}\,\langle
\Omega|\hat{\rho}_{k}|\Omega\rangle$. Except near slot borders the
$Q$-distribution before measurement is the (weighted) mixture of the
$Q$-distributions of the possible reduced states $\hat{\rho}_{k}%
$~\cite{Kofl2008}:\ $Q(\Omega)\approx%
{\textstyle\sum\nolimits_{k}}
w_{k}\,Q_{k}(\Omega)$.

The time evolution of the system obeys the Leggett-Garg inequality and
macrorealism, if the $Q$-distribution at any time $t_{j}$ without prior
measurements is approximately the same as the weighted mixture (over all
possible outcomes $k$) of the $Q$-distributions that resulted from an
intermediate measurement at any time $t_{i}$ ($t_{i}\!<\!t_{j}$) and then
evolved to $t_{j}$~\cite{Kofl2008}:%
\begin{equation}
Q(\Omega,t_{j})\approx%
{\displaystyle\sum\nolimits_{k}}
w_{k,t_{i}}\,Q_{k,t_{i}}(\Omega,t_{j})\label{eq Q cond}%
\end{equation}
This is the \textit{necessary and sufficient condition for macrorealism}, as
it incorporates both macrorealism per se and non-invasive measurability
(together with induction). If it is fulfilled, the Leggett-Garg inequality
follows~\cite{Kofl2008}. We will use this condition and not the Leggett-Garg
inequality to test macrorealism. Under non-classical Hamiltonians we
understand those that violate eq.~(\ref{eq Q cond}), and an example is (in
units where $\hbar=1$)%
\begin{equation}
\hat{H}=\text{i}\,\omega\left(  \left\vert -j\right\rangle \!\left\langle
+j\right\vert -\left\vert +j\right\rangle \!\left\langle -j\right\vert
\right)  .\label{eq Schroe}%
\end{equation}
Here, $\omega$ is the precession frequency and $\left\vert +j\right\rangle $
($\left\vert -j\right\rangle $) is the eigenstate of the spin-$j$ operator's
$z$-component with maximal (minimal) eigenvalue, or equivalently, the coherent
state pointing to the north (south). (For states of light, nonlinear media may
be used to implement similar interactions~\cite{Jeon2006}.) Starting from the
initial state along north, i.e.\ $|\Psi(0)\rangle=\left\vert +j\right\rangle
$, the Hamiltonian (\ref{eq Schroe}) produces an oscillating macroscopic
superposition state%
\begin{equation}
|\Psi(t)\rangle=\cos(\omega t)\left\vert +j\right\rangle +\sin(\omega
t)\left\vert -j\right\rangle .\label{eq psi}%
\end{equation}
The spin effectively behaves as a two-level system, albeit its two states are
macroscopically distinct. Given that the spin itself is isolated,
coarse-grained measurements or the fact that only measurement apparatuses
couple to an environment~\cite{Zure2003} do not prevent a violation of the
Leggett-Garg inequality and macrorealism~\cite{Kofl2008}. \textit{But what
happens if the system itself is continuously monitored by an environment?}

Given the Hamiltonian (\ref{eq Schroe}) with the time evolution operator
$\hat{U}_{t}=\exp(-$i$\hat{H}t)=\cos(\omega t)\left(  \left\vert
+j\right\rangle \!\left\langle +j\right\vert \!+\!\left\vert -j\right\rangle
\!\left\langle -j\right\vert \right)  +\sin(\omega t)\left(  \left\vert
-j\right\rangle \!\left\langle +j\right\vert \!-\!\left\vert +j\right\rangle
\!\left\langle -j\right\vert \right)  +%
{\textstyle\sum\nolimits_{m=-j+1}^{j-1}}
\left\vert m\right\rangle \!\left\langle m\right\vert $, let us approximate
the effects of (dephasing) system decoherence by the following simplified
model: The initial state along north, $\hat{\rho}(0)=\left\vert
+j\right\rangle \!\left\langle +j\right\vert $, freely evolves without
decoherence a short time $\Delta t$ to%
\begin{equation}
\hat{\rho}(\Delta t)=\cos^{2}(\omega\Delta t)\left\vert +j\right\rangle
\!\left\langle +j\right\vert +\sin^{2}(\omega\Delta t)\left\vert
-j\right\rangle \!\left\langle -j\right\vert +\text{c.t.}
\label{eq rho Delta t}%
\end{equation}
where the coherence terms \textquotedblleft c.t.\textquotedblright\ are of the
form $\left\vert +j\right\rangle \!\left\langle -j\right\vert $ and
$\left\vert -j\right\rangle \!\left\langle +j\right\vert $. Now we assume that
the macroscopic spin system (e.g.\ say $j\!\sim\!10^{23}$) decoheres very
rapidly (in the standard pointer basis of $\left\vert +j\right\rangle $ and
$\left\vert -j\right\rangle $), for instance due to the fact that a single
qubit from the environment couples to it in a controlled-not
interaction~\cite{Zure2003}, becomes inaccessible immediately afterwards, and
does not interact (recohere) with it anymore. If it is impossible to make
(joint) measurements on the environmental qubit (and our spin system), the
partial trace over the qubit of the total density matrix has to be performed,
which kills the coherence terms in eq.~(\ref{eq rho Delta t}), leading to the
decohered state of the system: $\hat{\rho}(\Delta t)=\cos^{2}(\omega\Delta
t)\left\vert +j\right\rangle \!\left\langle +j\right\vert +\sin^{2}%
(\omega\Delta t)\left\vert -j\right\rangle \!\left\langle -j\right\vert $.
Repeating the alternating sequence of free time evolution and rapid
decoherence, we obtain the general expression for the (decohered) state at
time $n\Delta t$:%
\begin{equation}
\hat{\rho}(n\Delta t)=A_{n}\left\vert +j\right\rangle \!\left\langle
+j\right\vert +(1\!-\!A_{n})\left\vert -j\right\rangle \!\left\langle
-j\right\vert . \label{eq rho n}%
\end{equation}
The survival probability to find the state along north, $A_{n}$, can be
retrieved from the recurrence relation $A_{n}=c\,A_{n-1}%
+(1\!-\!c)\,(1\!-\!A_{n-1})$ with integer $n$, $c\equiv\cos^{2}(\omega\Delta
t)$, and $A_{0}=1$. If $\Delta t$ is not too small (to avoid a quantum
Zeno-like freezing of the initial state~\cite{Misr1977} in this model) but
smaller than the dynamical timescale of the Hamiltonian, $\omega^{-1}$, the
probability $A_{n}$ decays to $A_{\infty}=\tfrac{1}{2}$ in a way which can be
very well approximated by%
\begin{equation}
A(t)=\tfrac{1}{2}\,(1+\text{e}^{-\nu t}) \label{eq exp decay}%
\end{equation}
with $\nu\approx2\sin^{2}(\omega\Delta t)/\Delta t$ the characteristic decay
rate. For times $t\gg\nu^{-1}$ the state asymptotically approaches an equal
weight statistical mixture $\hat{\rho}(\infty)=\tfrac{1}{2}\,(\left\vert
+j\right\rangle \!\left\langle +j\right\vert +\left\vert -j\right\rangle
\!\left\langle -j\right\vert )$. (In contrast, the probability to find the
state along north, i.e.\ $\left\vert +j\right\rangle $, at time $t$ if no
environmental decoherence takes place is given by $\cos^{2}(\omega t)$, where
this characteristic cosine-law allows to violate the Leggett-Garg inequality.)

As we will see below, the results of this simple model are generic for the
wide class of exact decoherence models with dephasing environments. The
environmental microscopic degrees of freedom drive the system into a mixture,
but it does not leave the subspace spanned by $\left\vert +j\right\rangle $
and $\left\vert -j\right\rangle $, and never populates any of the other spin
$z$-component eigenstates. The particular decay form (\ref{eq exp decay}) of
the survival probability is expected to hold very well in all cases where the
characteristic decoherence time $\tau_{\text{dec}}$, suppressing off-diagonal
elements in a density matrix, is fast compared to the dynamical timescale of
the Hamiltonian (\ref{eq Schroe}), i.e.\ whenever $\tau_{\text{dec}}\gg
\omega^{-1}$.

Let us now investigate what this means for the violation of the Leggett-Garg
inequality. We use eq.~(\ref{eq rho n}) in its continuous form, i.e.\ $\hat
{\rho}(t)=A(t)\left\vert +j\right\rangle \!\left\langle +j\right\vert
+[1\!-\!A(t)]\left\vert -j\right\rangle \!\left\langle -j\right\vert $, with
$A(t)$ the survival probability. If no (coarse-grained) measurement takes
place, the spin's $Q$-distribution at time $t_{j}$---i.e.\ the left-hand side
of eq.~(\ref{eq Q cond})---is given by%
\begin{equation}
Q(t_{j})=A(t_{j})\,Q_{\text{north}}+[1\!-\!A(t_{j})]\,Q_{\text{south}},
\label{eq lhs}%
\end{equation}
where $Q_{\text{north}}$ ($Q_{\text{south}}$) is the $Q$-distribution of a
spin pointing to the north (south). If a measurement takes place at the
intermediate time $t_{i}$ $(0<t_{i}<t_{j})$, the weighted mixture of the
reduced and evolved $Q$-distributions---i.e.\ the right-hand side of
eq.~(\ref{eq Q cond})---reads%
\begin{align}
&  \;\;\;\;\{A(t_{i})\,A(t_{j}\!-\!t_{i})+[1-A(t_{i})]\,[1-A(t_{j}%
\!-\!t_{i})]\}\,Q_{\text{north}}\nonumber\\
&  +\{A(t_{i})\,[1-A(t_{j}\!-\!t_{i})]+[1-A(t_{i})]\,A(t_{j}\!-\!t_{i}%
)\}\,Q_{\text{south}}. \label{eq rhs}%
\end{align}
Without loss of generality, we can set $A(t)=\tfrac{1}{2}\,[1+a(t)]$ with some
function $a(t)$ such that $A(t)$ is always between 0 and 1. We then find that
the non-invasiveness-condition (\ref{eq Q cond}), i.e.\ the equality of
(\ref{eq lhs}) and (\ref{eq rhs}), translates into the condition
$a(t_{j})=a(t_{i})\,a(t_{j}\!-\!t_{i})$. This is fulfilled \textit{if and only
if} $a(t)$ has the form e$^{-\nu t}$, where the solution with negative $\nu$
is excluded because $a(t)$ must always be between $-1$ and $+1$. This means
that the only allowed form of $A(t)$ is eq.~(\ref{eq exp decay}). Therefore,
in all decoherence models producing an exponential decay of the survival
probability---and only in those---the system's time evolution under the
Hamiltonian (\ref{eq Schroe}) fulfills the condition (\ref{eq Q cond}) for
non-invasive measurability, and consequently macrorealism is satisfied.

However, in the case of non-classical Hamiltonians, dephasing decoherence (and
therefore collapse models due to a universal noise background or gravitational
self energy which also have only a dephasing effect~\cite{Ghir1986})
\textit{cannot} account for a continuous spatiotemporal description of the
macroscopic spin variables. To see this, it is enough to use coarse-grained
measurements corresponding to only three different angular regions, one
covering the northern part, one the equatorial region, and one the southern
part. According to eq.~(\ref{eq rho n}), the initial spin along north can be
found pointing to the south at some later time, \textit{although it did not go
through the equatorial region}. No classical Hamilton function can achieve
such discontinuous \textquotedblleft jumps\textquotedblright\ of a spin vector.

The key point here is that in classical physics as well as in MR\&C we have
\textit{differential equations for observable quantities} such as spin
directions. Under all circumstances, these equations evolve the observables
\textit{continuously} through real space and time. In quantum mechanics,
however, the situation is very different. The Schrödinger equation evolves the
state vector continuously through Hilbert space and, under non-classical
Hamiltonians, one cannot give a continuous spatiotemporal description of the
coarse-grained (macroscopic) observables, even if macrorealism itself is
valid.\begin{figure}[t]
\begin{center}
\includegraphics[width=0.45\textwidth]{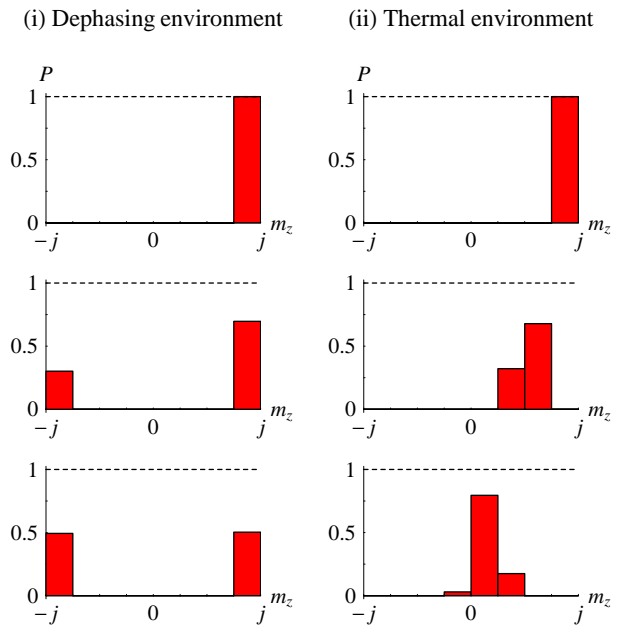}
\end{center}
\par
\vspace{-1.5cm}\caption{(Color online.) Snapshots of the probability
distribution $P$ that the magnetization along $z$ direction, $m_{z}$, takes
values in the indicated intervals from south ($-j$) to north ($+j$). The
system Hamiltonian is given by eq.~(\ref{eq system}). Left column, dephasing
environment: $P$ is computed at the successive dimensionless times $\omega
t=0,5,15$, respectively (from top to bottom). Intermediate values of $m_{z}$
are never possible and no continuous spatiotemporal description can explain
the shift in probability from north to south without populating values in
between. Right column, strong thermal environment: The plots are at $\omega
t=0,1,3$, respectively. The evolution is continuous and a continuous
spatiotemporal description in terms of classical laws of motions can exist.}%
\label{Figure2}%
\end{figure}

In the last part, we will investigate two decoherence models in detail, using
numerical solutions of the Lindblad master equation \cite{Lind1976,Breu2001}:%
\begin{equation}
{\frac{\text{d}}{\text{d}t}\,}\hat{\rho}(t)=-\text{i\thinspace}[\hat{H}%
,\hat{\rho}]-{\frac{1}{2}}\sum\nolimits_{k}\left(  [\hat{L}_{k}\hat{\rho}%
,\hat{L}_{k}^{\dag}]+[\hat{L}_{k},\hat{\rho}\hat{L}_{k}^{\dag}]\right)  ,
\label{eq master}%
\end{equation}
with the Lindblad operators $\hat{L}_{k}$. We assume the spin-$j$ system to
consist of $N=2j$ spin-$\tfrac{1}{2}$ particles and consider%
\begin{equation}
\hat{H}=\tfrac{1}{2}{\,}\omega{\,}(\hat{\sigma}_{1}^{+}\hat{\sigma}_{2}%
^{+}\dots\hat{\sigma}_{N}^{+}-\hat{\sigma}_{1}^{-}\hat{\sigma}_{2}^{-}%
\dots\hat{\sigma}_{N}^{-}), \label{eq system}%
\end{equation}
which is, up to a constant factor, the Hamiltonian (\ref{eq Schroe}). Here,
$\hat{\sigma}_{k}^{\pm}\equiv\hat{\sigma}_{k}^{x}\pm\,$i$\,\hat{\sigma}%
_{k}^{y}$ is the combination of Pauli $x$ and $y$ spin operators. (i) In the
first case, similar to the scenario above, the Lindblad operator corresponds
to a \textit{local dephasing} \cite{Breu2001}: $\hat{L}_{\text{dp}}=%
{\textstyle\sum\nolimits_{i=1}^{N}}
\gamma_{\text{dp}}\,\hat{\sigma}_{i}^{+}\hat{\sigma}_{i}^{-}$. (ii) In the
second case, a \textit{thermal environment} is modeled by \cite{Breu2001}
$\hat{L}_{\text{th}}=\frac{1}{2}\sum\nolimits_{i=1}^{N}\gamma_{\text{th}}%
{\,}[(\bar{n}\!+\!1)\,\hat{\sigma}_{i}^{-}-\bar{n}\,\hat{\sigma}_{i}^{+}]$.
The coupling parameter for the case of dephasing is set to $\gamma_{\text{dp}%
}=1$ so that the Lindblad and the Hamiltonian part in the master equation are
of the same order. The average number of excitations $\bar{n}$ in the thermal
environment is proportional to the temperature. We considered the case when
$\bar{n}\gg1$ and the coupling is such that $\gamma_{\text{th}}\bar{n}=1$. The
evolutions for the two cases were computed up to the times that are needed to
approximately reach the corresponding stationary distributions. For the
reasons of numerical convenience, the master equation (\ref{eq master}) was
solved using solutions of an equivalent stochastic nonlinear Schrödinger
equation for the quantum trajectories in the system's Hilbert space of pure
states. The particular form of the stochastic equation that we have used is
the one given by the theory of quantum state diffusion \cite{Perc1999}, but
other forms of equivalent stochastic equations or the direct solutions of
(\ref{eq master}) would give the same results. The evolution of the
expectations for an arbitrary observable is then given by averaging over many
stochastic trajectories ($10^{3}$ in our computation).

The results of the numerical computations are illustrated in the histograms in
Figure~\ref{Figure2}. We took $N=2j=10$ spin-$\tfrac{1}{2}$ particles,
initially all with spin along $z$, and computed the values of the
(dimensionless) magnetization along $z$, $\hat{m}_{z}=\tfrac{1}{2}%
{\textstyle\sum\nolimits_{i=1}^{N}}
\hat{\sigma}_{i}^{z}$. The possible domain of outcomes $m_{z}\in\lbrack-j,j]$
is divided into intervals and the probability $P$ that, without any
intermediate measurement, $m_{z}$ is in one of the intervals at a particular
time is shown. The two columns correspond to snapshots at successive moments
in the dimensionless time $\omega t$ with dephasing (left column) and thermal
decoherence (right column). Apparently, the evolution of $m_{z}$ under
decoherence by dephasing is \textit{discontinuous} in the sense that
intermediate intervals between north and south are never populated. As in the
simple model above, eq.~(\ref{eq rho n}), the environment just kills
off-diagonal terms in the density matrix. On the other hand, the strong
thermal environment actively perturbs the diagonal elements of the matrix and
produces a continuous evolution and therefore in principle allows a continuous
spatiotemporal description in terms of classical laws of
motion.\begin{figure}[t]
\begin{center}
\includegraphics[width=0.48\textwidth]{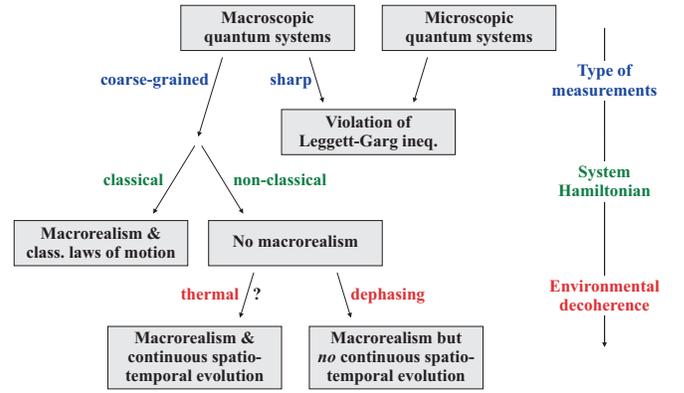}
\end{center}
\par
\vspace{-0.25cm}\caption{(Color online.) Microscopic systems as well as
macroscopic systems under sharp measurements allow to violate the Leggett-Garg
inequality \cite{Kofl2007,Kofl2008}. Since no macroscopic (coarse-grained)
\textquotedblleft classical\textquotedblright\ observables are involved, one
cannot speak about a violation of macrorealism in these cases. Under
coarse-grained measurements of a macroscopic system and classical Hamiltonians
not only macrorealism is valid but also classical laws of motion can emerge
\cite{Kofl2007}. Non-classical Hamiltonians---building up macroscopic
superpositions in time---allow to violate macrorealism even under
coarse-grained measurements \cite{Kofl2008}. In this work, we have shown that
decoherence then establishes macrorealism but that for dephasing environments
it cannot account for a continuous spatiotemporal description of the evolution
of macroscopic observables. Whether thermal environments can achieve such a
description in general remains an open question.}%
\label{Figure3}%
\end{figure}

\textit{Conclusion}.---Figure~\ref{Figure3} gives an overview regarding the
quantum-to-classical transition. The introduction of the concept
\textit{macrorealism and continuity} (MR\&C) in this work is at least of
twofold significance: First, it opens up the chance for experiments to
demonstrate a certain level of non-classicality despite the action of
decoherence. Second, it shows that the collapse models which have been put
forward to forbid macroscopic quantum effects \cite{Ghir1986} are insufficient
on their own, as---like dephasing---they only destroy superpositions but would
not in general ensure a continuous evolution of macroscopic properties.

This work was supported by the Austrian Science Foundation FWF (SFB, Project
No.\ P19570-N16, and CoQuS), the European Commission through Project QAP
(No.\ 015846), WUS Austria, ÖAD, and the Serbian Ministry of Science (Contract No.\ 141003).


\begin{thebibliography}{99}                                                                                               %


\bibitem {Legg1985}A. J. Leggett and A. Garg, Phys. Rev. Lett. \textbf{54},
857 (1985).

\bibitem {Zure2003}W. H. Zurek, Rev. Mod. Phys. \textbf{75}, 715 (2003); M.
Schlosshauer, Rev. Mod. Phys. \textbf{76}, 1267 (2004).

\bibitem {Legg2002}A. J. Leggett, J. Phys.: Cond. Mat. \textbf{14}, R415 (2002).

\bibitem {Schr1935}E. Schrödinger, Die Naturwissenschaften \textbf{48}, 807 (1935).

\bibitem {Arnd1999}M. Arndt \textit{et al}., Nature \textbf{401}, 680 (1999);
J. R. Friedman \textit{et al}., Nature \textbf{406}, 43 (2000); B. Julsgaard,
A. Kozhekin, and E. S. Polzik, Nature \textbf{413}, 400 (2001).

\bibitem {Ghir1986}G. C. Ghirardi, A. Rimini, and T. Weber, Phys. Rev. D
\textbf{34}, 470 (1986); P. Pearle, Phys. Rev. A \textbf{39}, 2277 (1989); R.
Penrose, Gen. Rel. Grav. \textbf{28}, 581 (1996).

\bibitem {Kofl2007}J. Kofler and \v{C}. Brukner, Phys. Rev. Lett. \textbf{99},
180403 (2007).

\bibitem {Kofl2008}J. Kofler and \v{C}. Brukner, Phys. Rev. Lett.
\textbf{101}, 090403 (2008).

\bibitem {Agar1981}G. S. Agarwal, Phys. Rev. A \textbf{24}, 2889 (1981); G. S.
Agarwal, Phys. Rev. A \textbf{47}, 4608 (1993).

\bibitem {Radc1971}J. M. Radcliffe, J. Phys. A: Gen. Phys. \textbf{4}, 313
(1971); P. W. Atkins and J. C. Dobson, Proc. R. Soc. A \textbf{321}, 321 (1971).

\bibitem {Jeon2006}H. Jeong \textit{et al}., Phys. Rev. A \textbf{70},
061801(R) (2004); H. Jeong, M. Paternostro, and T. C. Ralph, Phys. Rev. Lett.
\textbf{102}, 060403 (2009).

\bibitem {Misr1977}B. Misra and E. C. G. Sudarshan, J. Math. Phys.
\textbf{18}, 756 (1977).

\bibitem {Lind1976}{G. Lindblad, Commun. Math. Phys. \textbf{48}, 119 (1976).}

\bibitem {Breu2001}{H.-P. Breuer and F. Petruccione, \textit{The Theory of
Open Quantum Systems} (Oxford University Press, Oxford, 2001).}

\bibitem {Perc1999}{I. C.\ Percival, \textit{Quantum State Diffusion}
(Cambridge University Press, Cambridge, 1999).}
\end{thebibliography}
\end{document}